\documentclass[prb,aps,twocolumn,amsmath,amssymb,floatfix,showpacs,
superscriptaddress]{revtex4}

\usepackage[dvips]{graphics}
\usepackage{color}
\definecolor{gold}{rgb}{0.85,0.66,0}
\definecolor{dgreen}{rgb}{0.3,0.6,0.2}

\begin{document}

\title{\textcolor{dgreen}{Integer quantum Hall effect in a square 
lattice revisited}}

\author{Santanu K. Maiti}

\email{santanu@post.tau.ac.il}

\affiliation{School of Chemistry, Tel Aviv University, Ramat-Aviv,
Tel Aviv-69978, Israel}

\author{Moumita Dey}

\affiliation{Theoretical Condensed Matter Physics Division, Saha
Institute of Nuclear Physics, Sector-I, Block-AF, Bidhannagar,
Kolkata-700 064, India}

\author{S. N. Karmakar}

\affiliation{Theoretical Condensed Matter Physics Division, Saha
Institute of Nuclear Physics, Sector-I, Block-AF, Bidhannagar,
Kolkata-700 064, India}

\begin{abstract}

We investigate the phenomenon of integer quantum Hall effect in a 
square lattice, subjected to a perpendicular magnetic field, through 
Landauer-B\"{u}ttiker formalism within the tight-binding framework.
The oscillating nature of longitudinal resistance and near 
complete suppression of momentum relaxation processes are examined
by studying the flow of charge current using Landauer-Keldysh 
prescription. Our analysis for the lattice model corroborates the 
finding obtained in the continuum model and provides a simple physical 
understanding.

\end{abstract}


\maketitle

\section{Introduction}

The phenomenon of the integer quantum Hall effect (IQHE) is one of the
most significant discoveries in condensed matter physics~\cite{klit}.
At much low temperatures, typically below $1\,$K, this phenomenon occurs
in two-dimensional ($2$D) electron systems in presence of a strong 
perpendicular magnetic field. It has been observed that the Hall 
resistance is quantized in units of $h/2e^2$ with a great accuracy,
specified in parts per million, and due to this impressive accuracy 
it is utilized as a resistance standard~\cite{hart}. Such a precise 
quantization of the Hall resistance comes from almost complete 
suppression of momentum relaxation processes in the quantum Hall 
regime. In a finite width conductor at high magnetic fields, the 
states carrying current in one direction are spatially separated 
from those carrying currents in the opposite direction, and as a 
result the overlaps between these two groups of states get 
significantly reduced which results in a suppression of backscattering. 
The near complete suppression of backscattering between the forward 
and backward propagating states leads to a truly ballistic conductor 
even in the presence of impurities. The quantized nature of Hall 
resistance looks like the quantized resistance of conventional ballistic 
conductors, but in these conductors precise quantization cannot be 
achieved since the backscattering processes are not fully 
removed~\cite{datta}. 

It has also been noticed that in a Hall bar the longitudinal resistance 
oscillates as a function of the Fermi energy $E_F$. Naturally it may 
occur to our mind that when the Fermi energy matches with anyone of the
Landau levels i.e., with a peak in the density of states (DOS) profile, 
the resistance should have a minimum. But the real fact is completely 
opposite to that. The resistance becomes minimum whenever the Fermi 
energy lies between two Landau levels where the density of states also 
has a minimum, and, it leads to an immense curiosity and gives rise to 
the question
about the current carrying states through the sample. It has been 
observed that the states those are situated near the edges of the sample,
the so-called {\em edge states}, play an important role for carrying 
current when the longitudinal resistance has a minimum~\cite{halp,
macd}. It is quite surprising that at the minima, the longitudinal 
resistance is almost zero and the electrons are able to traverse
a long distance without dissipating their momentum. It emphasizes that 
something special must be taking place which provide near complete 
suppression of the momentum relaxation processes in this regime. 

Several studies already exist in the literature which deal with the 
quantized nature of the Hall resistance, oscillating behavior of the 
longitudinal resistance, existence of the edge 
\begin{figure}[ht]
{\centering \resizebox*{7cm}{3.25cm}{\includegraphics{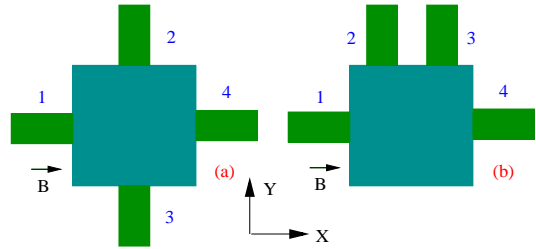}}\par}
\caption{(Color online). Four probe set-ups for determining (a) the Hall
resistance and (b) the longitudinal resistance.}
\label{model}
\end{figure}
states and almost complete suppression of the backscattering processes 
in a Hall bar based on the {\em continuum} model~\cite{lau,thou1,thou2,
pran,avr,klit1}. However, no rigorous effort has been made so far, to 
the best of our knowledge, to unravel all these features within a 
{\em discrete lattice} model. In the present work we consider a 
two-dimensional square lattice in presence of a perpendicular magnetic 
field and establish them describing the system by the tight-binding 
(TB) model. Using the Landauer-B\"{u}ttiker formalism~\cite{datta} 
we find the longitudinal and transverse conductivities. On the other 
hand, to investigate the behavior of the current carrying states when 
Fermi energy lies between two Landau levels or coincides with a Landau 
level we use Landauer-Keldysh prescription~\cite{niko1,niko2}. From 
this study we can clearly describe the momentum relaxation processes. 
The physical picture about QHE that emerges from our present study based
on the discrete lattice model of the Hall bar is exactly the same as 
obtained in the continuum model and provides much insight to the problem 
under realistic condition. This approach could be much more
useful in further investigation of the unconventional quantum Hall effect in 
graphene where the Hall conductivity is half-integer quantized~\cite{novo,kim},
showing plateaus at $\sigma_{xy}=(4e^2/h)(n+1/2)$, $n=0$, $\pm 1$, $\pm 2$,
$\dots$, since this lattice model might be adapted with more convenience
in graphene and other topologically insulating materials~\cite{hata,long}. 
It is our first step towards this direction.

In what follows, we present the results. In Section $2$, we describe the 
model and theoretical formulation. Section $3$ contains the numerical 
results and related  discussions, and we draw our conclusions in Section 
$4$.

\section{The model and Theoretical formulation}

The quantum Hall system is shown schematically in Fig.~\ref{model}, where 
a square lattice, placed in a magnetic field $\vec{B}$ perpendicular to its 
plane, is attached to four finite width leads. The TB Hamiltonian for a 
square lattice with $N$ atomic sites in both the $x$ and $y$ directions 
reads,
\begin{eqnarray}
H &=& \sum_{m,n} \epsilon_{m,n} c_{m,n}^{\dag} c_{m,n} + t \sum_{m,n} 
\left(c_{m+1,n}^{\dag}c_{m,n} \right. \nonumber \\
& + & \left. c_{m,n+1}^{\dag}c_{m,n}\,e^{i \theta_m} + \mbox{h.c.}
\right).
\label{equ1}
\end{eqnarray}
Here $\epsilon_{m,n}$ represents the site energy of an electron at the 
lattice site ($m$,$n$), $m$ and $n$ being the $x$ and $y$ co-ordinates 
of the site, respectively. $c_{m,n}^{\dagger}$ ($c_{m,n}$) corresponds 
to the creation (annihilation) operator of an electron at the site 
($m$,$n$) and $t$ is the nearest-neighbor hopping integral. The phase factor 
$\theta_m=2\pi m B a^2$ ($a$ being the lattice spacing) arises due to the
magnetic field \mbox{\boldmath$B$} with the choice of the vector potential 
\mbox{\boldmath$A$}$\,=$($0, Bx, 0$).

The side-attached leads are taken as semi-infinite square lattice ribbons,
which are also described by TB Hamiltonians similar to Eq.~\ref{equ1},
where the hopping terms do not invoke the phase factor $e^{i\theta_m}$
since $\vec{B}=0$ in the leads. The lead Hamiltonians are parametrized 
by constant on-site potential $\epsilon_0$ and nearest-neighbor hopping 
strength $t_0$. The hopping integrals between the boundary sites of the 
leads and the quantum Hall system are parametrized by $\tau$. The number
of sites in the transverse direction of the leads is denoted by $M_y$.

In order to find the Hall resistance using Landauer-B\"{u}ttiker formalism,
we connect the leads to the sample as prescribed in Fig.~\ref{model}(a) and 
restrict ourselves within the coherent transport regime~\cite{datta}. The
voltage leads, namely, lead-2 and lead-3 are attached on the opposite sides
of the sample. In this formalism one can treat all the leads (i.e., current 
and voltage leads) on equal footing and express the current in the lead-p 
as $I_p=\sum_{q=1}^4 G_{pq}\left[V_p-V_q\right]$, $G_{pq}$ being the 
conductance coefficient related to transmission coefficient $T_{pq}$ 
between the lead-p and lead-q by the expression $G_{pq}=(2e^2/h) T_{pq}$
(factor $2$ accounts for the spin of the electrons). Here $V_p$ is the 
applied bias voltage in the lead-p. Since the currents in the various 
leads depend only on voltage differences among them, we can set one of 
the voltages to zero without loss of generality. Here we set $V_4=0$. 
It allows us to write the currents in three other leads in the form 
of matrix equation, 
\mbox{\boldmath{$V$}}={\boldmath{$\mathcal{G}$}}$^{-1}$\mbox{\boldmath{$I$}},
where \mbox{\boldmath{$\mathcal{G}$}} is a $3\times 3$ matrix whose
elements are expressed as follows: $\mathcal{G}_{ii}=\sum_k^{k\ne i}G_{ik}$
and $\mathcal{G}_{ij}=-G_{ij}$, where $k$ runs from $1$ to $4$. The above 
matrix equation simplifies in the form:
\begin{equation}
\left\{\begin{array}{c}
V_1 \\
V_2 \\
V_3 \\
\end{array} \right\} 
= 
\left\{\begin{array}{ccc}
R_{11} & R_{12} & R_{13} \\
R_{21} & R_{22} & R_{23} \\
R_{31} & R_{32} & R_{33} \\
\end{array} \right\}
\left\{\begin{array}{c}
I_1 \\
I_2 \\
I_3 \\
\end{array} \right\}
\label{equ4}
\end{equation}
where, the matrix \mbox{\boldmath{$R$}} is the inverse of 
\mbox{\boldmath{$\mathcal{G}$}}. In our four-terminal set-up 
(Fig.~\ref{model}(a)) the Hall resistance is determined from the relation,
\begin{equation}
R_H=\left[\frac{V_2-V_3}{I_1}\right]_{I_2=I_3=0}=R_{21}-R_{31}.
\label{equ5}
\end{equation}

On the other hand, to determine the longitudinal resistance we connect 
the leads to the conductor as shown in Fig.~\ref{model}(b), attaching the 
voltage leads lead-2 and lead-3 on the same side of the conductor. For 
this set-up also the above expressions remain invariant, and the 
longitudinal resistance is given by $R_L=R_{21}-R_{31}$. Obviously, for 
this set-up the numerical values of $R_{21}$ and $R_{31}$ are different 
from those of the previous set-up (Fig.~\ref{model}(a)).

With these $R_L$ and $R_H$ we can determine the longitudinal and Hall 
conductivities where the conductivity tensor $\sigma_{\mu\nu}$ is 
the inverse of the resistivity tensor $\rho_{\mu\nu}$. Mathematically
it is expressed as $\sigma_{\mu\nu}=\left(\rho_{\mu\nu}\right)^{-1}$,
in which $\rho_{xx}$ and $\rho_{xy}$ correspond to the longitudinal 
and Hall resistivities, respectively. For a square lattice, 
$\rho_{xx}=\rho_{yy}=R_L$ and $\rho_{xy}=-\rho_{yx}=R_H$. Therefore,
we can write the final expressions of the conductivities as~\cite{hajdu},
\begin{equation}
\sigma_{xx}=\sigma_{yy}=\frac{R_L}{R_L^2+R_H^2}; \,
\sigma_{yx}=-\sigma_{xy}=\frac{R_H}{R_L^2+R_H^2}.
\label{equ7}
\end{equation}

Next, we find the bond charge current between the atomic sites $n$ and
$n^{\prime}$ by the Landauer-Keldysh method using the 
relation~\cite{niko1,niko2},
\begin{equation}
\langle \hat{J}_{nn^{\prime}}\rangle = \frac{et}{h} 
\int\limits_{E_F-eV_0/2}^{E_F+eV_0/2} \mbox{Tr}_s
\left[G_{nn^{\prime}}^<(E) - G_{n^{\prime}n}^<(E) \right]dE.
\label{equ10}
\end{equation}
It actually describes the flow of charges which start at the site $n$ 
and end up at the site $n^{\prime}$. In this expression $G^<$ is the 
conventional lesser Green's function~\cite{niko1} and $V_0$ is the 
applied potential difference between the biased leads. Here trace Tr$_s$ 
is performed in the spin Hilbert space, writing the Hamiltonian 
Eq.~\ref{equ1} including the spin of the electrons.

Throughout the numerical calculations we set $\epsilon_0=0$, lattice 
spacing $a=1$ and fix the hopping integrals ($t$, $t_0$ and $\tau$) at 
$1$. All energies are measured in units of $t$ and we choose $c=e=h=1$. 

\section{Numerical results and discussion}

\begin{figure}[ht]
{\centering \resizebox*{4.25cm}{3.75cm}{\includegraphics{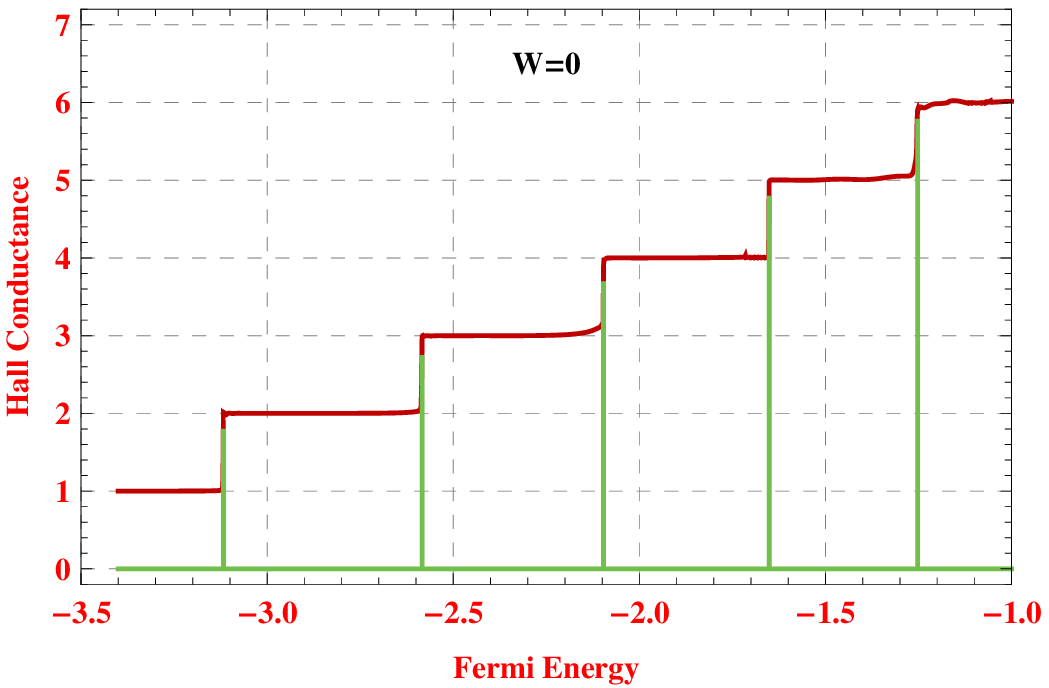}}
\centering \resizebox*{4.25cm}{3.75cm}{\includegraphics{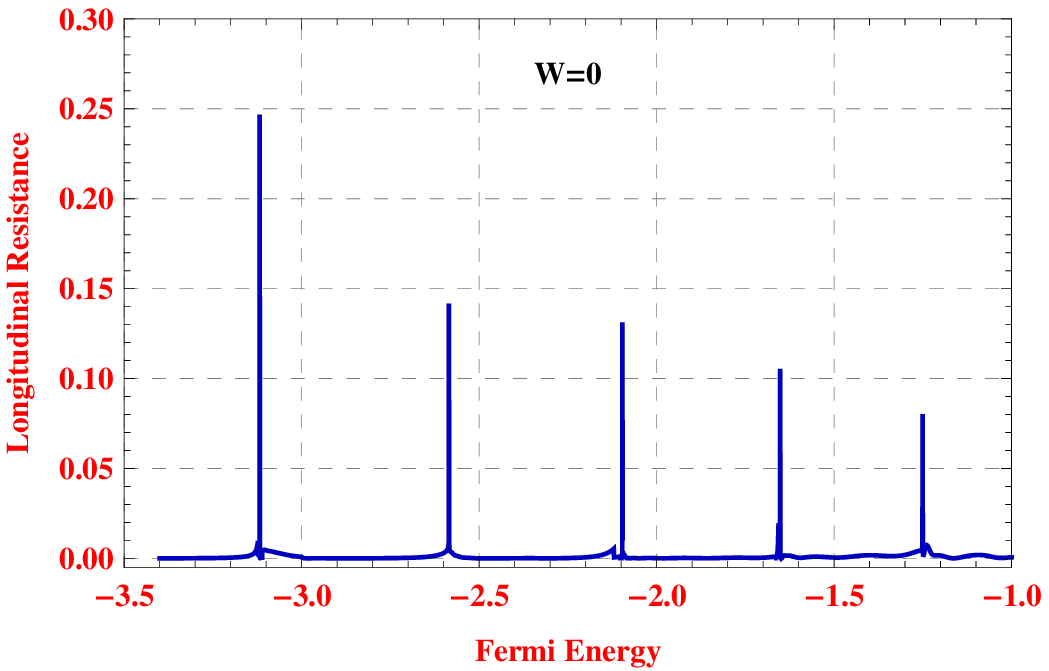}}
\par}
\caption{(Color online). Left panel: Hall conductance (red line) as a
function of Fermi energy and the electronic density of states (green line)
for a perfect ($W=0$) square lattice of size $20\times 20$ considering
$B=0.05$. Right panel: Longitudinal resistance as a function of Fermi
energy for the same parameter values.}
\label{result1}
\end{figure}

In the left panel of Fig.~\ref{result1} we show the quantized nature of 
the Hall conductance together with the density of states for an ordered 
($W=0$, $W$ measures the disorder strength in the sample) square lattice. 
The sharp peaks in the DOS are associated with the Landau bands those are 
produced in the presence of magnetic field $B$. For a fixed sample size, 
the total number of Landau bands and the number of states in each band
strongly depend on the strength of the applied magnetic field. Here we
choose the uniform magnetic field $B$ in the form $1/Q$, $Q$ being an 
integer, and it generates $Q$ number of Landau bands in the energy 
spectrum. This choice of the magnetic field makes each Landau band 
populated by $N^2/Q$ states for a square lattice of size $N\times N$. 
Thus, for the set of parameter values chosen in Fig.~\ref{result1}, we will
get twenty Landau bands at the peaks of the DOS spectrum and each band 
accommodates twenty states. Since here we present the results only for a 
particular energy range, few Landau bands are visible.
From the spectrum it is observed that the Hall conductance increases 
in discrete steps whenever the Fermi energy crosses peaks in the DOS 
profile. This is what we would expect from an ordinary ballistic waveguide 
where two-terminal conductance changes in integer steps associated with 
the integer number of sub-bands or transverse modes at the Fermi energy 
$E_F$~\cite{datta}. On the other hand, plateaus appear in the Hall 
conductance when the Fermi energy lies between two Landau bands. At 
these plateaus the Hall conductance gets the value $(2e^2/h) M$, where 
$M$ corresponds to the total number of Landau bands below the Fermi energy.
This is exactly equal to the number of edge states at the Fermi energy
and these states play the equivalent role similar to the transverse
modes in a conventional ballistic waveguide. 

\begin{figure}[ht]
{\centering \resizebox*{4.25cm}{3.75cm}{\includegraphics{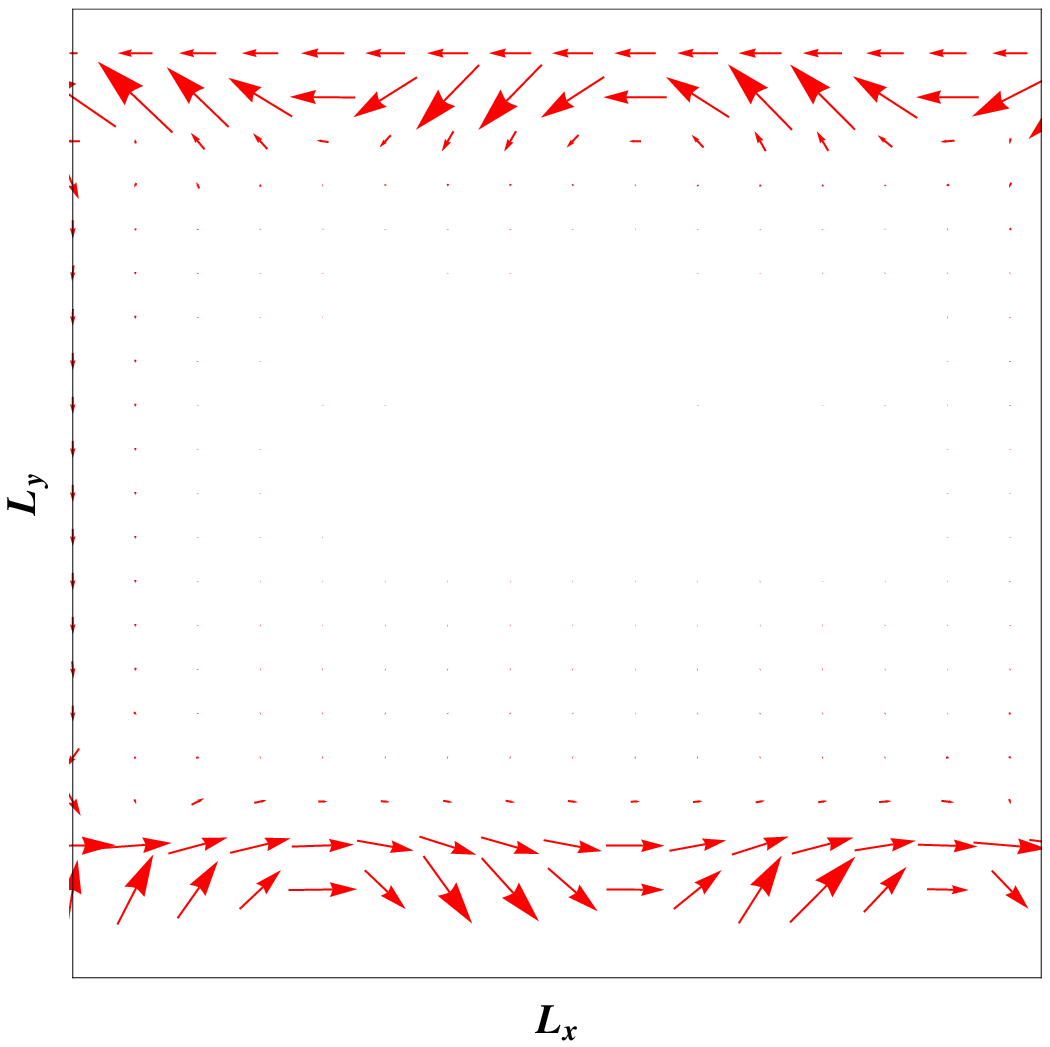}}
\centering \resizebox*{4.25cm}{3.75cm}{\includegraphics{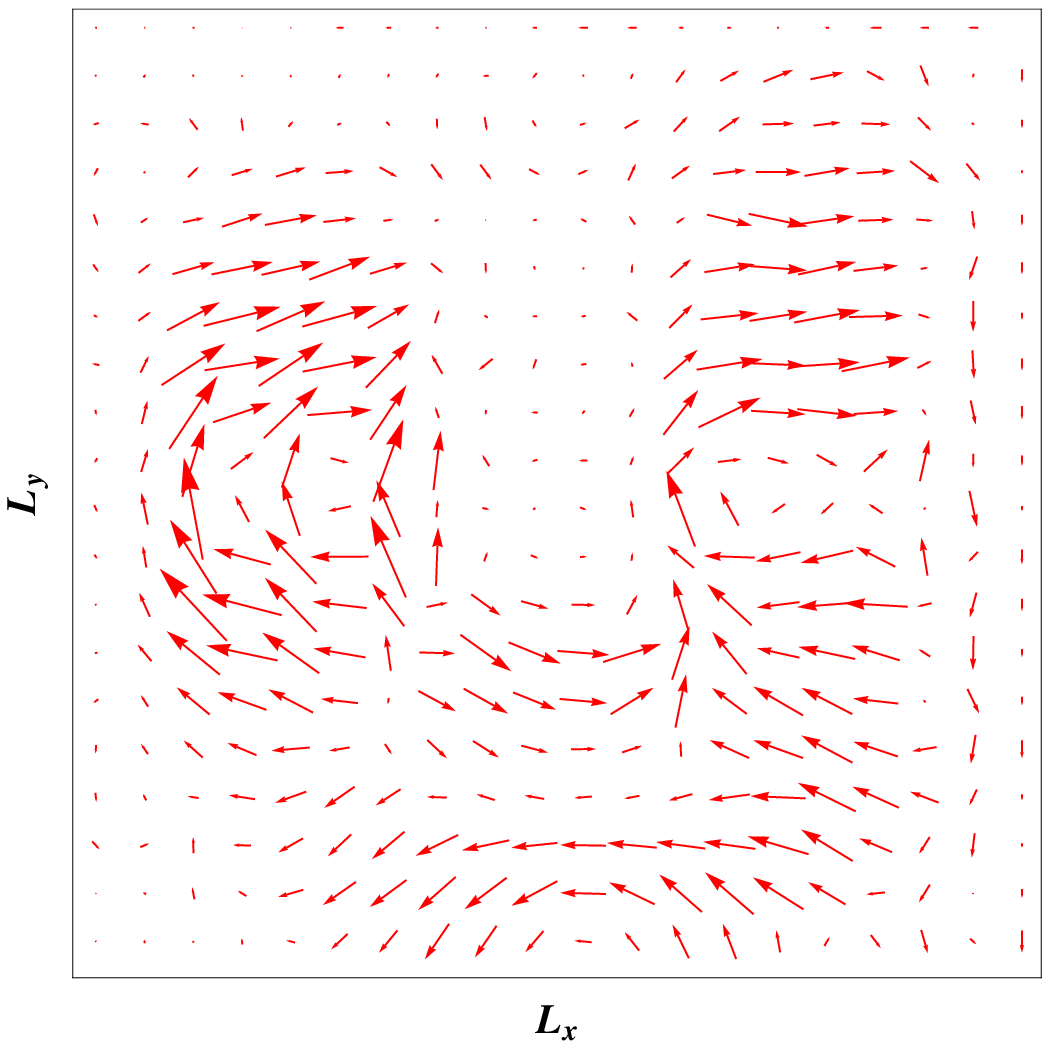}}
\par}
\caption{(Color online). Flow of charge current in a perfect ($W=0$)
square lattice of size $20 \times 20$ considering $B=0.1$, where the
left and right panels correspond to $E_F=-2.8$ and $-1.46$, respectively.
Left panel: Charge current flows only through the edges and no current is
obtained in the bulk. Right panel: Charge current flows through the bulk.}
\label{result2}
\end{figure}

Though the quantized Hall conductance resembles to the quantized 
conductance in traditional ballistic waveguides, but the strange precise 
quantization can never be achieved in ordinary ballistic conductors since 
the backscattering processes are not fully suppressed. The almost complete 
elimination of the backscattering processes can only be achieved in the 
quantum Hall regime which provides surprisingly precise quantized Hall 
conductance.
To reveal this fact, we examine the flow of charge current in a perfect 
conductor when the Fermi energy lies within a plateau region. The result 
is shown in the left panel of Fig.~\ref{result2}, where $E_F$ is set at 
$-2.8$, and it predicts that the charge current flows only through the 
edges of the conductor and no current is available in the bulk.
Very nicely we notice that the states carrying current from the left side 
of the sample to the right one are spatially separated from 
those carrying current in the reverse direction. As a result the overlap 
between the forward and backward propagating states is almost reduced to 
zero leading to near complete suppression of the backscattering processes 
and thereby the momentum relaxation process. In the continuum model 
these states are referred as the positive and negative $k$-states and they 
are commonly named as the {\em edge states}. At high magnetic field 
we get practically zero overlap between these states. Due to this complete
elimination of momentum relaxation, the edge states carrying current 
towards the right side of the sample are in equilibrium with the left
lead, while the other states carrying current in the opposite direction
are in equilibrium with the right lead. Therefore, the difference in
voltage measured by two voltage probes placed anywhere on the same
side (the so-called longitudinal voltage) of the conductor gets zero, 
while a non-zero value of the Hall voltage is obtained when it is 
measured by two voltage probes placed anywhere on the opposite sides 
of the conductor. 

The almost entire suppression of the momentum relaxation process, when 
Fermi energy lies in the plateau regions, gives rise to essentially zero 
longitudinal resistance exactly what we have obtained numerically (see 
the right panel of Fig.~\ref{result1}). The resistance of a conductor 
is governed by the rate at which the electrons can loose their momentum. 
To relax the momentum an electron has to be scattered from the left side 
to the right side of the conductor through the allowed energy eigenstates 
in the bulk of the conductor. This is practically zero (see the left panel 
of Fig.~\ref{result2}) as long as the Fermi energy lies between two Landau 
levels. The non-zero value of the longitudinal resistance is obtained 
exclusively when the backscattering process takes place and it would be 
maximum when the Fermi energy synchronizes with a Landau band (right panel
of Fig.~\ref{result1}). To articulate the authenticity, in the right panel
of Fig.~\ref{result2}, we expose the nature of bond charge current in the
conductor when the Fermi energy is fixed at a Landau band ($E_F=-1.46$).
\begin{figure}[ht]
{\centering \resizebox*{4.25cm}{3.75cm}{\includegraphics{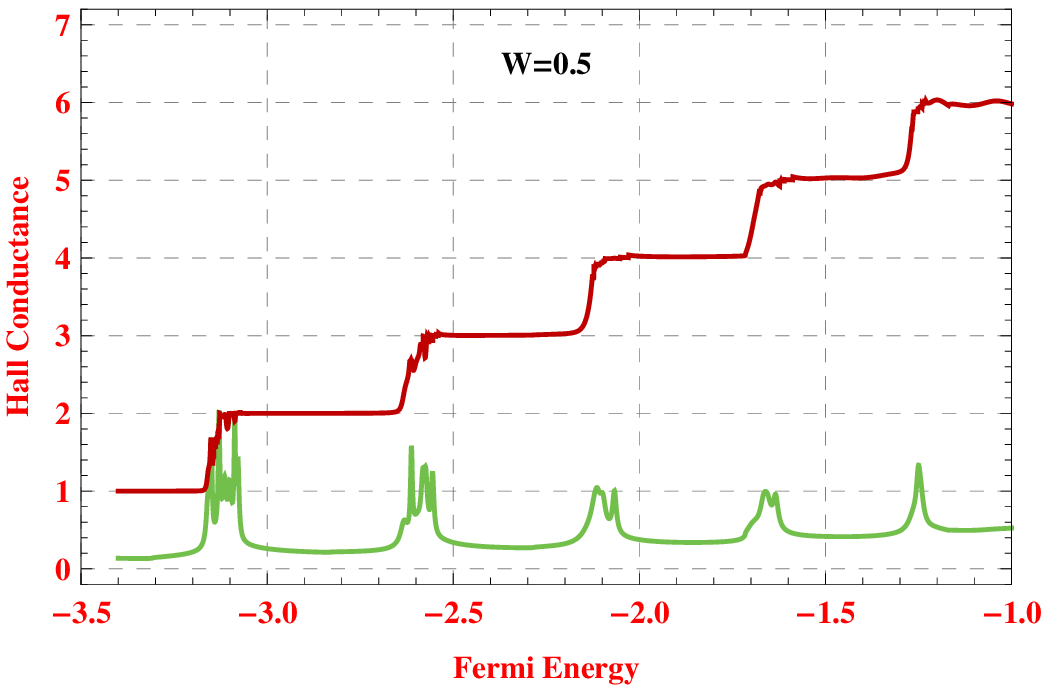}}
\centering \resizebox*{4.25cm}{3.75cm}{\includegraphics{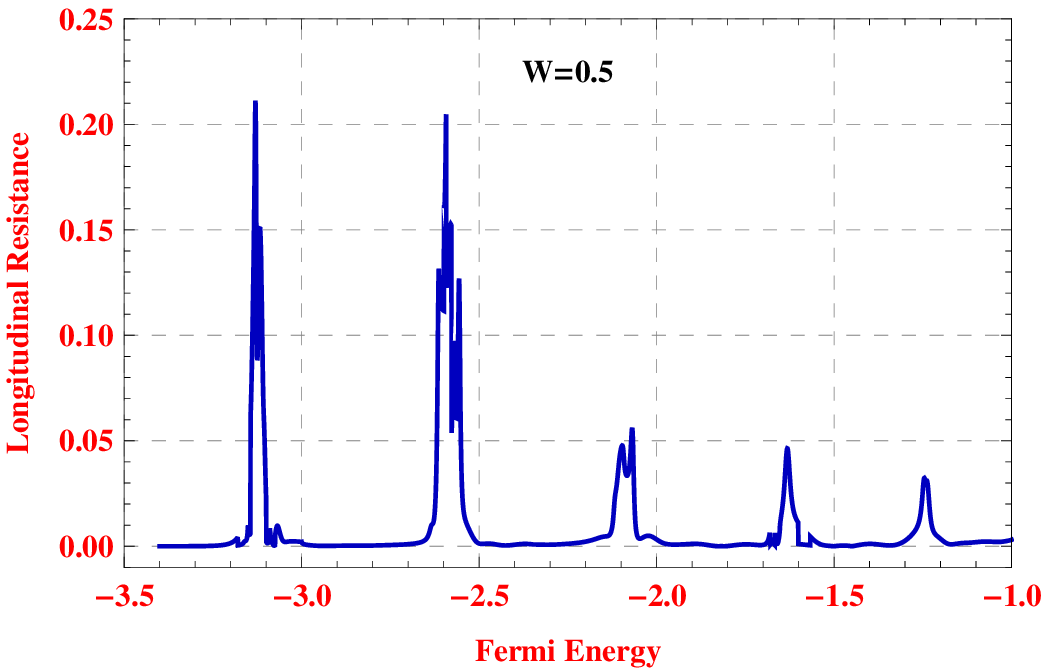}}\par}
\caption{(Color online). Left panel: Hall conductance (red line) as a
function of Fermi energy and the electronic density of states (green
line) for a disordered ($W=0.5$) square lattice of size $20\times 20$
considering $B=0.05$. Right panel: Longitudinal resistance as a function
of Fermi energy for the same parameter values.}
\label{result3}
\end{figure}
It predicts a continuous flow of charge form one edge to the other, and
certainly, electrons can scatter from the left to the right side of the 
conductor through its bulk. This scattering leads to the longitudinal
resistance. Thus, each time the Fermi energy crosses a Landau band, 
we get maximum backscattering and hence a peak in the longitudinal 
resistance (see the right panel of Fig.~\ref{result1}).

\begin{figure}[ht]
{\centering \resizebox*{4.25cm}{3.75cm}
{\includegraphics{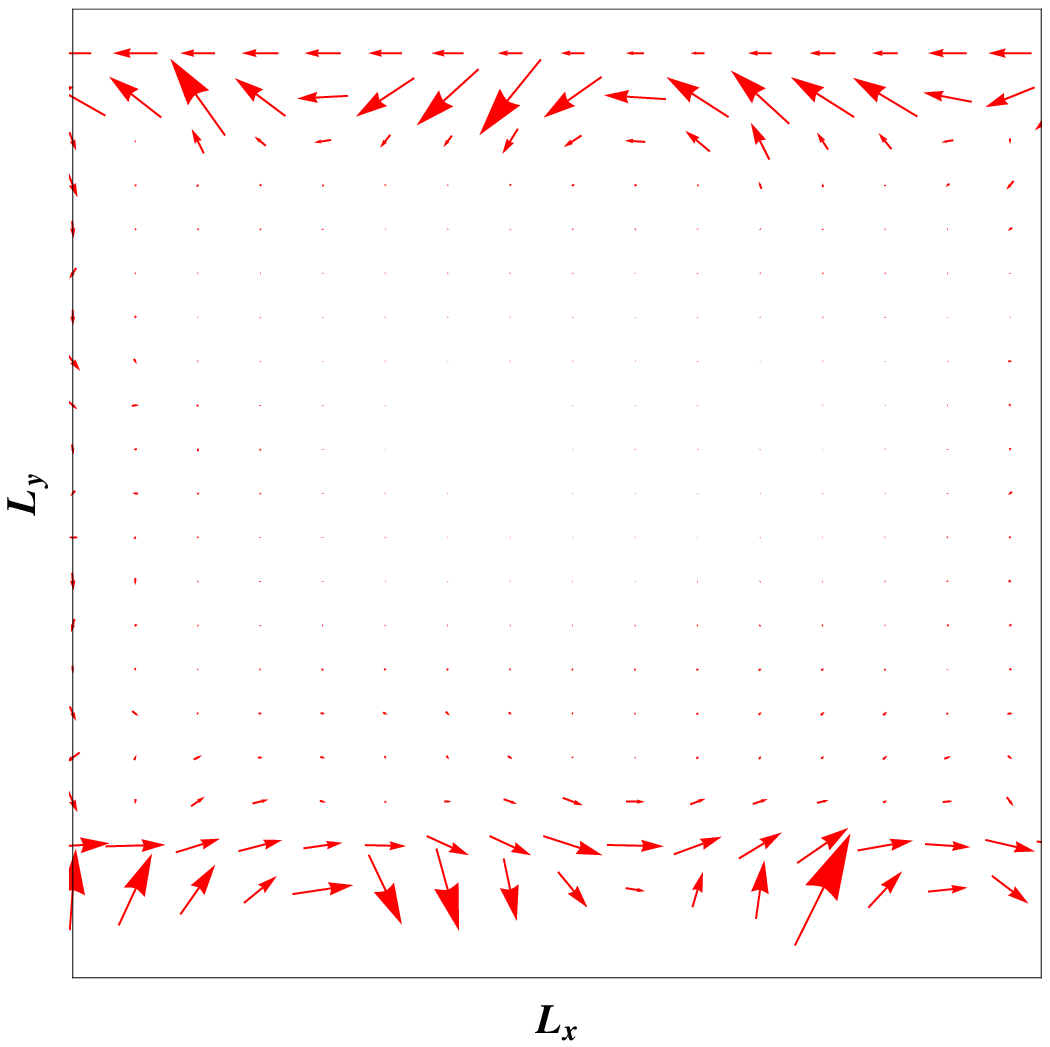}}
\centering \resizebox*{4.25cm}{3.75cm}
{\includegraphics{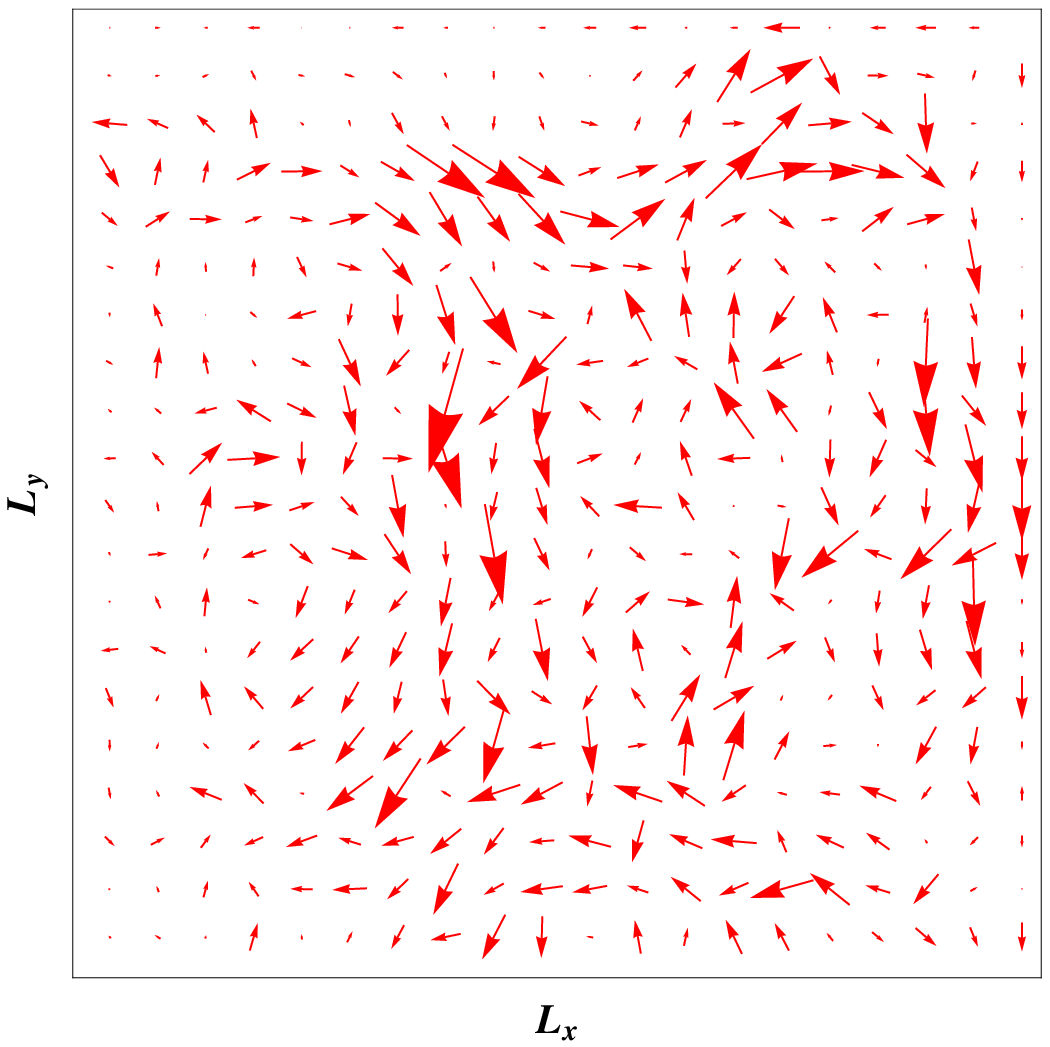}}\par}
\caption{(Color online). Flow of charge current in a disordered ($W=1$) 
square lattice of size $20 \times 20$ considering $B=0.1$, where the 
left and right panels correspond to $E_F=-2.8$ and $-1.46$, respectively.
Left panel: Charge current flows only through the edges and no current is 
obtained in the bulk. Right panel: Charge current flows through the bulk.}
\label{result4}
\end{figure}

We now present the results for a disordered square lattice of size 
$20\times 20$ in Fig.~\ref{result3}. Disorder is introduced via a random 
distribution (width $W=0.5$) of values of the on-site potentials (diagonal 
disorder), and results averaged over hundred disorder configurations are
presented. In the presence of disorder sharp Landau levels are broadened
(green line in the left panel of Fig.~\ref{result3}), but it does not 
affect the quantized nature of the Hall conductance. However, in contrast
to the perfect conductor, the sharpness of the Hall conductance gets
reduced as observed in the left panel of Fig.~\ref{result3} (red curve). 
At the plateau regions the Hall conductance is again precisely quantized 
in units of $2e^2/h$ and such conductance quantization, even in the 
presence of impurities, comes due to the complete spatial separation 
of the forward and backward propagating states which becomes apparent in 
Fig.~\ref{result4} (left panel). On the other hand, when the Fermi 
energy lies on a bulk Landau band, electrons can easily scatter through 
the interior of the conductor, giving rise to non-zero bond charge 
currents throughout the conductor (right panel of Fig.~\ref{result4}). 
This scattering yields the longitudinal resistance (see the right panel 
of Fig.~\ref{result3}). 

It is important to note that with the increase of the disorder 
strength Landau bands get broadened more by virtue of disorder. As a 
result of disorder localized states are appeared in the tails of the 
broadened Landau bands and they do not contribute to the transport. But, 
as we get the non-vanishing Hall conductance, all the states within the 
Landau bands cannot be localized. At the band centers extended states 
exist which carry the Hall current and the amount of current which is 
lost due to the formation of localized states in the tail of a Landau 
band is exactly compensated by the remaining extended 
states~\cite{prange,ando}. Since the localized states within the Landau 
band do not contribute to the current the Hall plateaus become broadened 
which has been clearly explained by Kramer {\em et al.}~\cite{kramer}. It 
is also to be noted that for large enough disorder strength and for weak 
magnetic field, quantum Hall plateaus gradually vanish from the higher 
energy side. We confirm it numerically. However, it was not the main 
motivation of our present work, and the disappearance of the integer 
quantum Hall effect due to disorder has already been reported in the 
literature~\cite{sheng1,sheng2}.

Before we end this section, we would like to point 
out that the disappearance of edge states and the appearance of finite
resistance associated with the backscattering processes, as a result
of movement of the Fermi level $E_F$, strongly depend on the system
size, magnetic field and also on the impurity strength. Here we describe
very briefly about the nature of quantum phase transition associated 
with the above mentioned factors, to make the present communication a self
contained study. The near zero value of the longitudinal resistance is
obtained only when the Fermi energy lies anywhere within the plateau 
regions and in this situation the backscattering processes are almost 
suppressed which result edge currents in the sample. While, for all 
other choices of $E_F$, the finite value of longitudinal resistance is
obtained i.e., the backscattering processes are dominated which provide 
bulk current. For a finite system size and fixed disorder strength, the 
appearance of distinct Landau levels, and hence the quantized nature of
Hall plateaus, depends on the choice of the magnetic field. If the field 
is too weak, then the Hall plateaus will no longer appear and the 
backscattering processes are also strong enough to produce bulk current. 
On the other hand, if we fix the system size and choose the magnetic field 
to a moderate one then the appearance of the edge currents or bulk current 
upon the movement of the Fermi level is significantly controlled by the 
impurity strength. For our system size ($20 \times 20$) we examine that 
when $W > 2$, the Hall plateaus almost disappear and we get bulk current 
for any choice of $E_F$ and this feature would be visible for much less
value of $W$ in the thermodynamic limit. Thus, we can emphasize that the 
quantum phase 
transition in the phenomenon of integer quantum Hall effect in a square 
lattice geometry strongly depends on the system size, disorder strength 
and the applied magnetic field. For a particular disorder strength and 
in presence of a moderate magnetic field, a functional dependence of the 
Fermi energy $E_F$ with system size may be obtained for the crossover 
between the edge currents and bulk currents. Since at this stage it is 
really too complicated to compute it numerically we can try to establish 
this functional form in our future work, but, we believe, the main essence 
of all these features are quite well understood from our present study.

\section{Conclusion}

In conclusion, we have investigated the phenomenon of integer quantum 
Hall effect in a discrete lattice model through Landauer-B\"{u}ttiker 
formalism. The almost complete suppression of momentum relaxation 
processes and the oscillating nature of longitudinal resistance are 
clearly explored by examining the flow of charge current using 
Landauer-Keldysh prescription. The results presented in this communication 
are worked out for absolute zero temperature. However, they should remain 
valid even in a certain range of finite temperatures ($\sim 300$\,K). This 
is because the broadening of the energy levels of the conductor due to the 
conductor-lead coupling is, in general, much larger than that of the 
thermal broadening~\cite{datta}. Our present study about QHE based on the 
discrete lattice model provides much better insight to the problem under 
realistic condition.

\end{document}